\documentclass[twocolumn,trackchanges]{aastex701}

\def\h{\hspace{-2.0 mm}}
\def\nd{\nodata}
\def\asec{$^{\prime\prime}$}

\begin{document}

\title{A Catalog of Mid-infrared Variable Sources in the Ecliptic Poles} 

\author[0000-0002-3560-0781]{Minjin Kim}
\affiliation{Department of Astronomy, Yonsei University, 
50 Yonsei-ro, Seodaemun-gu, Seoul 03722, Republic of Korea}
\email[show]{mkim.astro@yonsei.ac.kr}

\author[0000-0002-5346-0567]{Suyeon Son}
\affiliation{Kavli Institute for Astronomy and Astrophysics, 
Peking University, Beijing 100871, People's Republic of China}
\email[]{}

\author[0009-0009-9529-514X]{Shinyu Kim}
\affiliation{Department of Astronomy, Yonsei University, 
50 Yonsei-ro, Seodaemun-gu, Seoul 03722, Republic of Korea}
\affiliation{Department of Astronomy and Atmospheric Sciences, 
Kyungpook National University, Daegu 41566, Republic of Korea}
\email[]{}

\author[0000-0001-6947-5846]{Luis C. Ho}
\affiliation{Kavli Institute for Astronomy and Astrophysics, Peking University, Beijing 100871, People's Republic of China}
\affiliation{Department of Astronomy, School of Physics, Peking University, Beijing 100871, People's Republic of China}
\email{lho.pku@gmail.com}

\author[0000-0002-2770-808X]{Woong-Seob Jeong}
\affiliation{Korea Astronomy and Space Science Institute (KASI) 
776, Daedeok-daero, Yuseong-gu, Daejeon 34055, Republic of Korea}
\email[]{}

\author[0000-0003-1954-5046]{Bomee Lee}
\affiliation{Korea Astronomy and Space Science Institute (KASI) 
776, Daedeok-daero, Yuseong-gu, Daejeon 34055, Republic of Korea}
\email[]{}

\author[0000-0003-3078-2763]{Yujin Yang}
\affiliation{Korea Astronomy and Space Science Institute (KASI) 
776, Daedeok-daero, Yuseong-gu, Daejeon 34055, Republic of Korea}
\email[]{}


\begin{abstract}

We construct a catalog of mid-infrared (MIR) variable sources using the multi-epoch 3.6 (W1) and 4.5 $\mu$m (W2) dataset from the Near-Earth Object Wide-field Infrared Survey Explorer (NEOWISE) at the north and south ecliptic poles (NEP and SEP). The catalog provides well-sampled light curves that cover areas within a radius of 5 degrees from the poles, which are frequently observed by current and forthcoming missions. By carefully processing the NEOWISE data to secure reliable photometric measurements, we identified 2764 and 27581 variables in the NEP and SEP, respectively, using the probability deviating from the non-variable and the correlation coefficient between W1 and W2. Cross-correlation with various complementary datasets reveals that, in the NEP, variability is dominated by active galactic nuclei, whereas stellar objects are more common in the SEP due to its proximity to the Large Magellanic Cloud. In particular, proper motion measurements from Gaia and MIR color-color diagrams are ideal for narrowing down the physical origin of the MIR variable sources. We identify three MIR transients in the NEP. Interestingly, all coincide with obscured QSOs, suggesting a physical connection between transient events and circumnuclear obscuration. Finally, we discuss the potential applications of our catalog in synergy with existing and future time-domain surveys.            

\end{abstract}

\keywords{\uat{Active galactic nuclei}{16}}


\section{Introduction} 
While the brightness of most astronomical objects remains constant with time, occasional flux variations provide critical insights into the physical properties of variable objects \cite[e.g.,][]{herbst_1994, gautschy_1995, ulrich_1997}. Excluding transient events and eclipses, variability is most frequently observed in asteroids, young stellar objects (YSOs), evolved stellar populations, and compact objects such as neutron stars and black holes. These variations serve as a diagnostic tool; for instance, the timescale of variation acts as a proxy for the physical size of the light-emitting region or probes the physical origin of the variability and internal structure of stars \cite[e.g.,][]{conroy_2018, burke_2021, son_2025, kim_2026}. While variability has been extensively studied in the optical and radio regimes due to the abundance of time-series data \cite[e.g.,][]{hughes_1992, alcock_1997, macleod_2010, arevalo_2024}, other wavelengths, such as X-ray, UV, and infrared regimes, remain comparatively underexplored, which can be investigated with various current and upcoming telescopes \cite[e.g.,][]{frostig_2024}. 

Specifically, variability in the infrared (IR) regime is poorly understood due to a historical scarcity of time-domain data, despite its vital role in tracing dust emission and being relatively unaffected by dust extinction \cite[e.g.,][]{matsunaga_2011, sanchez_2017, green_2024}. Over the last decade, mid-infrared (MIR) variability studies have gained momentum following the Wide-field Infrared Survey Explorer (WISE) mission (\citealp{wright_2010}). Since 2010, WISE has conducted all-sky surveys with a cadence of approximately six months. Following the depletion of its cryogen, the mission continued as the NEOWISE project \citep{neowise}, utilizing its two ``warm'' detectors ($3.4\ \mu\text{m}$ and $4.6\ \mu\text{m}$) until mid-2024 (\citealp{mainzer_2011}). This mission has provided a unique, decadal multi-epoch dataset for the entire sky, enabling the systematic study of MIR variability across diverse populations of astronomical objects \cite[e.g.,][]{chen_2018, park_2021, son_2022, son_2023, aravindan_2024, kang_2025}. Using this dataset, long-term MIR variability and transients, such as tidal disruption events (TDEs) and AGN flares, have been extensively studied \cite[e.g.,][]{stern_2018, jiang_2021, son_2022b, wang_2022,  meisner_2023, paz_2024, masterson_2024, necker_2025}.

Despite the advantages of the WISE dataset, its relatively sparse cadence can limit the scientific robustness of derived physical properties. However, due to the sun-synchronous orbit of the WISE satellite, the north and south ecliptic poles (NEP and SEP, respectively) were sampled much more frequently than the rest of the sky. These regions offer exceptionally dense light curves, making them essential for high-cadence variability studies. Furthermore, the ecliptic poles are primary targets for current and upcoming missions, including SPHEREx (\citealp{bock_2026}), 7-dimensional survey (7DS; \citealp{kimj_2024}), and Legacy Survey of Space and Time (LSST; \citealp{ivezic_2019}). SPHEREx, in particular, will provide high-cadence MIR spectral data in these fields. Consequently, identifying MIR variables at the ecliptic poles is vital for two primary reasons. First, they serve as foundational reference sources for future high-cadence missions. Second, combining these variables with complementary multi-wavelength, multi-epoch datasets provides a unique opportunity to explore the physical properties and governing physics of variable sources. Motivated by these factors, this study generates a comprehensive catalog of MIR variables in the NEP and SEP utilizing multi-epoch WISE datasets. Throughout this paper, we adopt Vega magnitudes, unless otherwise noted. 


\section{Sample and Data} \label{sec:sample}
To generate catalogs of sources exhibiting MIR variability in the ecliptic poles, we initially collect MIR sources located within a radius of 5 degrees from the NEP and SEP. We require this parent sample to have a signal-to-noise ratio (S/N) greater than 10 in both the W1 and W2 bands in the combined dataset from AllWISE \citep{allwise} to ensure the detection of variability with reliable photometric accuracy. The selection area is designed to sufficiently cover the deep regions of SPHEREx for the NEP. However, for the SEP, the center of the SPHEREx deep region is shifted relative to WISE by $\sim8^{\circ}$ to avoid interference from the LMC. To maintain consistency, we applied the same area restrictions (a radius of $5^{\circ}$ from the SEP) as used for the NEP.
This results in $\sim0.6$ and $\sim1.0$ million objects in the NEP and SEP, respectively. The sample size is significantly larger for the SEP because it partially covers the Large Magellanic Cloud (LMC), leading to the inclusion of a substantial number of stellar objects.

While multi-epoch data from both AllWISE and NEOWISE are available for MIR variability studies, systematic offsets in the photometry between the two datasets are frequently observed (\citealp{mainzer_2014}). These offsets are often non-trivial to correct using simple methods and can result in the spurious detection of variability. To mitigate this risk, we choose to utilize only the NEOWISE dataset for our parent sample. To ensure the reliability of the photometric measurements retrieved from NEOWISE, we apply the following constraints: \texttt{qual\_frame} $>$ 0, \texttt{qi\_fact} $>$ 0, \texttt{saa\_sep} $>$ 0, \texttt{moon\_masked} = `00', \texttt{cc\_flags} = `0000', \texttt{w1rchi2} $< 10$, and \texttt{w2rchi2} $< 10$ \citep{son_2022}. These flagging criteria are specifically designed to grant reliable photometry and to minimize the presence of outliers in the photometry. Additionally, we use a matching radius of 1 \asec\ as the photometric outliers are often associated with a large spatial offset. 

\begin{figure*}[tp!]
\centering
\includegraphics[width=0.49\textwidth]{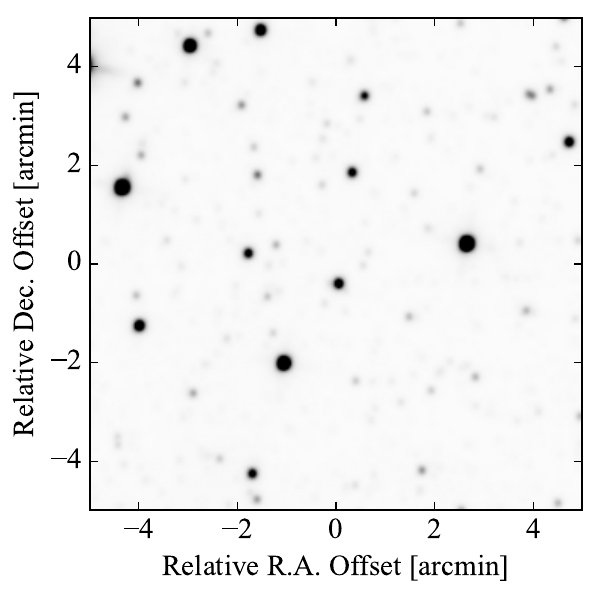}
\includegraphics[width=0.49\textwidth]{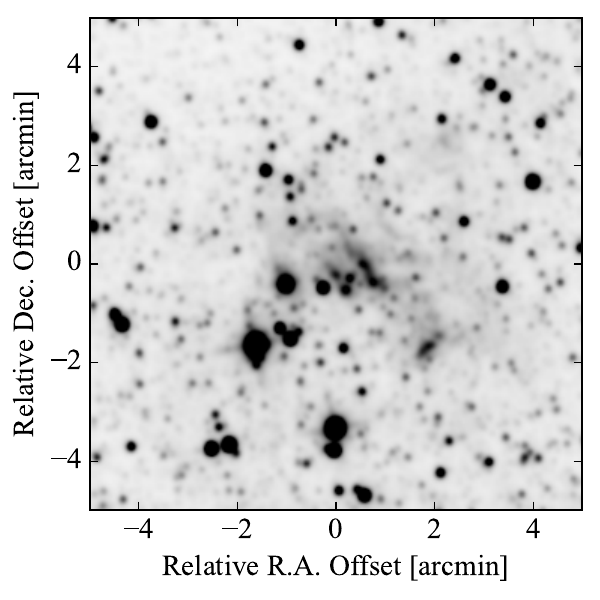}
\vspace{-0.3cm}
\caption{Representative cutouts of an ordinary area (left) and high-density stellar regions (right) within the SEP.}
\end{figure*}

Since a specific field is observed multiple times during each visit of the WISE mission survey, it is necessary to bin the multi-epoch data. To preserve the temporal resolution provided by the ecliptic poles, we adopt a 30-day bin size.
During the binning process, we apply $3\sigma$ clipping to effectively remove outliers, followed by averaging the remaining observed values within each bin. To ensure photometric reliability, we only utilize binned data if a given bin contains more than five individual observations. Furthermore, we restrict our study to targets with binned photometric data spanning more than 10 distinct epochs, a criterion that is crucial for robustly validating long-term variability. This leaves 4166016 and 785262 objects in the NEP and SEP, respectively.

In the NEOWISE mission, orbital decay caused a gradual change in the detector temperature. This, combined with seasonal temperature variations, led to fluctuations in the zero-point (e.g., \citealp{son_2026, kims_2026}). These effects are more pronounced in the W2 data, where the zero-point varies from $\sim -0.01$ to $\sim0.05$ mag, compared to the W1 data, which ranges between $\sim -0.01$ and $\sim0.01$ mag. Consequently, we correct for these variations using time-series temperature data from the focal plane\footnote{\url{https://irsa.ipac.caltech.edu/data/WISE/docs/release/NEOWISE/expsup/sec4_2d.html}}. Since the temperature data are available from modified Julian dates (MJD) 56784.3 to 60532.3, we restrict our analysis to WISE data within this time interval, which covers the entire NEOWISE dataset.   

\begin{figure}[tp!]
\centering
\includegraphics[width=0.49\textwidth]{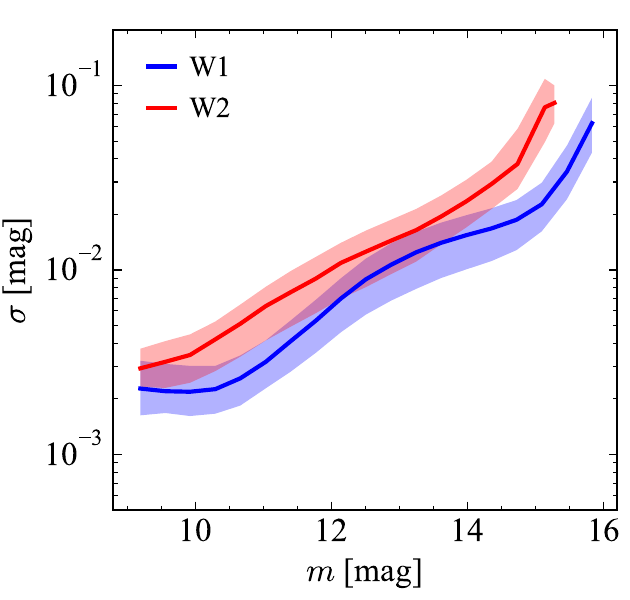}
\vspace{-0.3cm}
\caption{Median photometric errors as a function of magnitude for the W1 (blue) and W2 (red) bands. The shaded regions represent the 16th to 84th percentile range within each magnitude bin.}
\end{figure}

\begin{figure*}[tp!]
\centering
\includegraphics[width=0.49\textwidth]{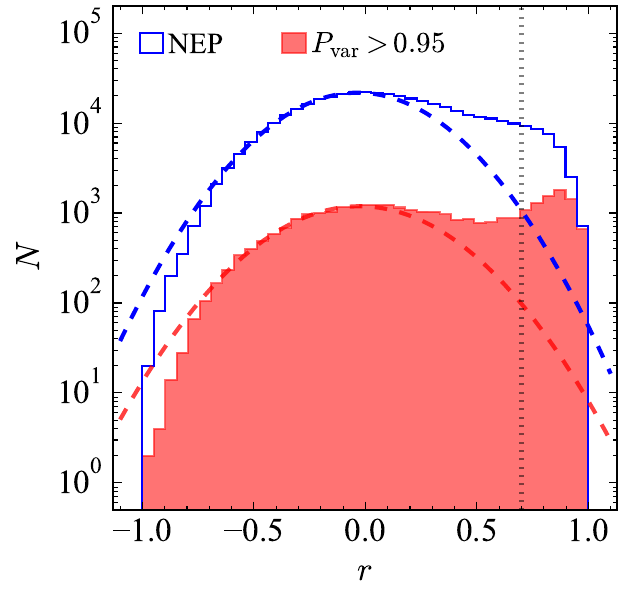}
\includegraphics[width=0.49\textwidth]{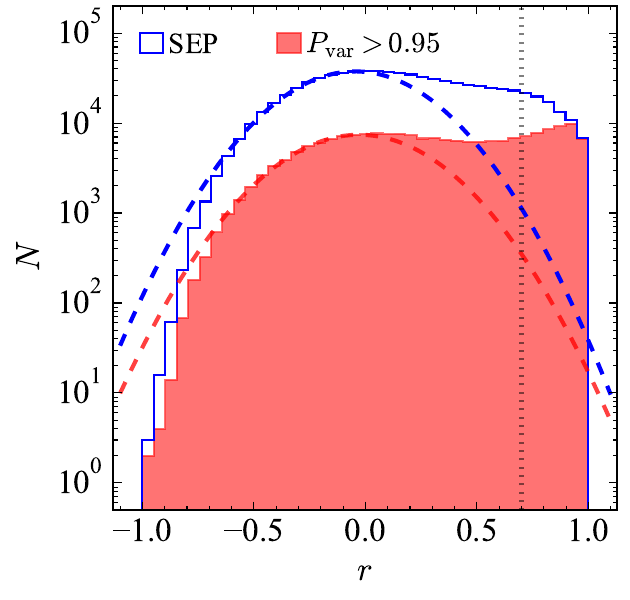}
\vspace{-0.3cm}
\caption{Distributions of the correlation coefficient ($r$) for the NEP (left) and SEP (right). While blue histograms represent the entire AllWISE sample with S/N $> 10$ in both W1 and W2 bands, sources satisfying $P_{\rm var} > 0.95$ for both bands are denoted by red-filled histograms. Dashed lines indicate Gaussian fits for sources with $r \leq 0$, and dotted lines represent our selection criteria (i.e., $r=0.7$) for variable sources.}
\end{figure*}

In general, when accounting only for extragalactic sources, source confusion is likely not severe in our dataset, given the adopted S/N cuts \cite[e.g.,][]{kimd_2024,huai_2025}. However, contributions from the stellar component can enhance blending. The SEP, in particular, suffers from this issue due to its proximity to the LMC. Figure 1 compares an ordinary area in the SEP with the dense regions of the LMC, clearly illustrating that confusion becomes problematic near the LMC. To quantify this, we cross-match the variable sources with the Gaia catalogs using a matching radius of 2\asec\ (see \S{4.2}). We find that only $2.3\%$ of sources have multiple Gaia counterparts in the NEP, whereas $\sim28.7\%$ exhibit multiple counterparts in the SEP. While a detailed discussion of blending is beyond the scope of this study, photometric data in the SEP near the LMC should be used with caution. 

While measurement errors for individual observations are provided by NEOWISE, photometric uncertainties in binned data are heavily influenced by the specific binning method employed. Consequently, we calculate these uncertainties based on the distribution of standard deviations ($\sigma$) derived from the light curves of individual targets. We find that $\sigma$ is strongly dependent on target magnitude in both the W1 and W2 bands. To represent the photometric uncertainty at a given brightness, we adopt the median value of $\sigma$ within each magnitude bin. This approach accounts for sources with intrinsic variability, which appear as outliers in the $\sigma$ distribution and could lead to a significant overestimation of uncertainties if an average were used. By using the median instead of the mean, we minimize the impact of these variables and ensure a more robust estimate of the noise. Finally, the photometric uncertainties for individual targets are derived for each epoch based on the empirically driven magnitude–error relation established from the entire sample (Fig. 2). For this calculation, we restrict our sample to the NEP, as the confusion issues in the SEP can introduce additional uncertainties. As there are insufficient samples to determine the photometric uncertainties at the bright end, we restrict the final sample to be fainter than 9 mag in both the W1 and W2 bands.

\section{Identification of Variable Sources}
To identify variable sources in the NEP and SEP using MIR light curves, we employ two parameters: (1) the probability that a source exhibits intrinsic variability based on the $\chi^2$ distribution ($P_{\rm var}$), and (2) the correlation coefficient ($r$) between the W1 and W2 bands. The parameter $P_{\rm var}$ quantifies the likelihood that a light curve deviates from a non-variable model. We define $\chi^2$ as $\displaystyle
\sum_{i=1}^{N}\frac{(m_i-\bar{m})^2}{\sigma_i^2}$, where $m_i$ and $\sigma_i$ represent the observed magnitude and corresponding uncertainty in each epoch, $N$ is the number of binned data points, and $\bar{m}$ is the mean magnitude from the light curve. Then $P_{\rm var}$ is computed from the $\chi^2$ distribution with $N-1$ degrees of freedom \cite[e.g.,][]{mclaughlin_1996, sanchez_2017}. 
This probability is calculated independently for the W1 and W2 bands, yielding $P_{\rm var,,W1}$ and $P_{\rm var,W2}$, respectively. Sources with $P_{\rm var} > 0.95$ in both W1 and W2 bands were initially identified as variable \cite[][]{lanzuisi_2014, sanchez_2017, son_2022}. Although $P_{\rm var}$ has been widely used in previous studies to identify variable sources in various time-series datasets, relying on this metric alone can lead to misclassification, as it is sensitive to outliers and can be significantly affected by inaccurately estimated measurement uncertainties.

\begin{figure*}[tp!]
\centering
\includegraphics[width=0.98\textwidth]{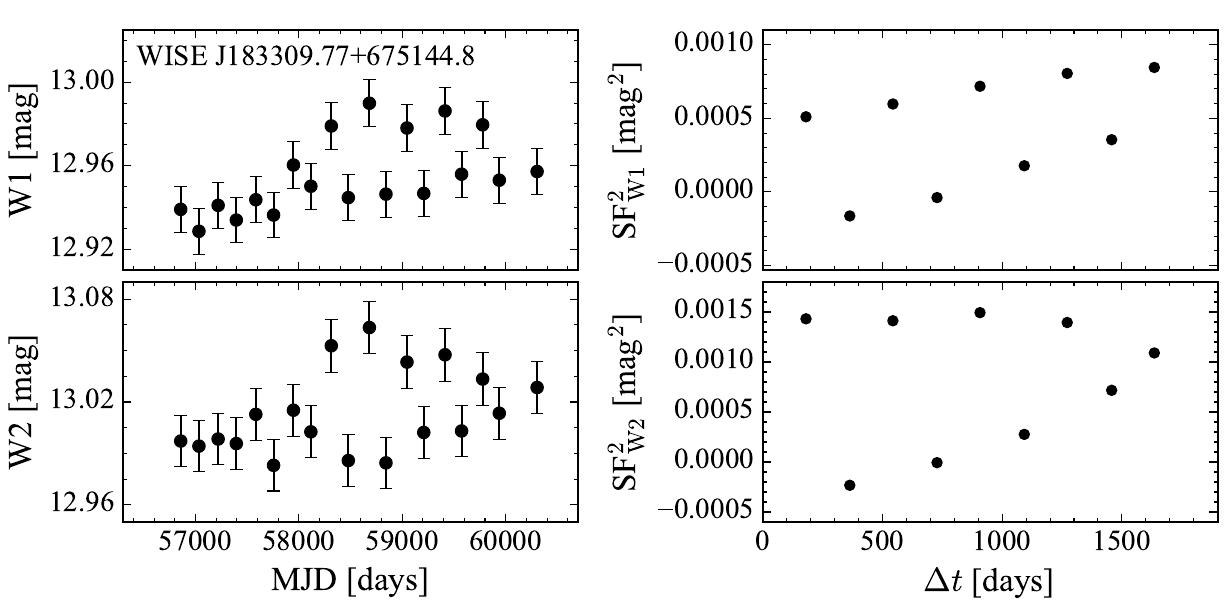}
\caption{Examples of W1 (top) and W2 (bottom) light curves for spurious sources affected by the afterimage of a nearby bright object. These exhibit quasi-periodic light curves with a period of $\sim 1\ {\rm yr}$. The corresponding structure functions are shown in the right panels.}
\end{figure*}

To enhance the robustness of our variability selection against systematics affecting $P_{\rm var}$, we additionally incorporate the correlation coefficient between the W1 and W2 bands \cite[e.g.,][]{2025ApJ...991...52A}. Variability driven by random photometric errors or poorly constrained uncertainties is expected to produce weak or no correlation between the two bands, resulting in low $r$ values. In contrast, intrinsic variability should manifest as coherent, correlated variations in both W1 and W2. As a result, the inclusion of $r$ provides a complementary and more systematics-resistant criterion for confirming genuine variability. This approach has been successfully employed in previous studies to identify AGNs. In this work, we adopt the Pearson correlation coefficient as the definition of $r$. 

The distribution of $r$ for the full sample is illustrated in Figure 3. A Gaussian fit applied to the $r \leq 0$ data suggests that while the distribution generally follows the Gaussian distribution, there is a clear excess at higher $r$ values. This indicates that $r$ is an effective metric for identifying variable sources. Although defining a precise threshold for variability is complex, the sample density increases significantly above $r \sim 0.7$. To ensure high sample purity and minimize noise from questionable sources, we define $r > 0.7$ as our selection criterion. While this conservative threshold may exclude some true variables, it prioritizes the reliability of the final catalog. Applying these selection criteria ($P_{\rm var} > 0.95$ and $r > 0.7$) results in 7644 and 48132 objects in the NEP and SEP, respectively.

\subsection{Removal of Spurious Sources}
Visual inspection of the light curves reveals a significant fraction of suspicious targets exhibiting periodic variations of $\sim12$ months (consistent with twice the individual visit cadence; Fig. 4). This phenomenon has been reported in previous studies and is attributed to persistence effects (afterimages) from nearby bright sources \cite[e.g.,][]{jiang_2021}. Because the survey scan direction reverses every six months, this persistence occurs every two visits, resulting in artificial periodic variability. To mitigate these artifacts, we employ the structure function (SF) of the light curve. The SF characterizes the variability amplitude as a function of time lag and is defined as: $${\rm SF}^2 (\Delta t) = \frac{1}{N_{\Delta t}} \sum_{i=1}^{N_{\Delta t}} [m(t) - m(t+\Delta t)]^2 - \sigma(t)^2 - \sigma(t+\Delta t)^2,$$ where $N_{\Delta t}$ is the number of pairwise combinations at time lag $\Delta t$, and $m$ and $\sigma$ are the observed magnitude and corresponding uncertainty at a given epoch. 

For a damped random walk process, the SF increases at short timescales and flattens at longer lags in ordinary AGNs. In contrast, periodic variability produces a periodic SF. For instance, artificial variables with a 12-month period exhibit a maximum SF at 6 months and a minimum at 12 months.          
Motivated by this, we identify the artificial variables using a ratio of ${\rm SF}^2(0.5\ \rm yr)$ and ${\rm SF}^2(1\ \rm yr)$. More specifically, we empirically determine the following criteria to select these suspicious sources:
\begin{itemize}
\item $\frac{{\rm SF}^2(1\ \rm yr)+{\rm SF}^2(2\ \rm yr)}{{\rm SF}^2(0.5\ \rm yr)+{\rm SF}^2(1.5\ \rm yr)} < 0.2$ at W1 or W2 bands.
\item ${\rm SF}^2(0.5\ \rm yr) > -0.0001$ mag for both W1 and W2 bands
\end{itemize}
This reduces the number of variables by $\sim64$\% and $\sim43$\% for the NEP and SEP, respectively. The final sample consists of 2764 in the NEP and 27581 in the SEP. Table 1 summarizes the final variables for the ecliptic poles and their associated selection parameters. In addition, the structure functions within 2 years are listed in Table 2.

\section{Complementary Data}
\subsection{QSO Catalogs}
To investigate the physical origin and characteristics of the variability in our sample, we incorporate complementary data from external surveys. Given that the observed variability may originate from AGNs, we cross-match our sources with three distinct AGN catalogs. First, we utilize the Dark Energy Spectroscopic Instrument (DESI) Data Release 1 (DR1; \citealp{desi_2025}). Although DESI DR1 is limited to the northern hemisphere, it provides robust spectroscopic classifications. Within the NEP, we find $747$ variables with a DESI DR1 counterpart within a $2^{\prime\prime}$ matching radius. Of these, $451$ are spectroscopically confirmed as QSOs, while $295$ are classified as galaxies, suggesting they are likely to be Type 2 AGNs. Only one source is classified as a stellar object. It is interesting to note that $\sim 80\%$ of spectroscopically confirmed DESI QSOs are non-variable sources in the MIR. As these MIR non-variable QSOs tend to exhibit higher redshifts ($z=1.5\pm0.7$) and are fainter in the W1 band (W1$=16.2\pm0.7$ mag) than MIR variable QSOs ($z=0.7\pm0.4$ and W1$=14.9\pm0.7$ mag), such a large fraction of non-detections for MIR variability is likely due to the relatively low sensitivity and short baseline of the WISE dataset.     

\begin{deluxetable*}{cccccccccccc}[ht!]
\tablewidth{0pt}
\scriptsize
\tablecaption{Photometric Properties of Variable Sources \label{tab:tab1}}
\tablehead{
\colhead{Name} & \colhead{\h R.A.} & \colhead{\h Dec.} &
\colhead{\h $P_{\rm var, W1}$} & \colhead{\h $P_{\rm var, W2}$} & \colhead{\h $r$} &
\colhead{\h $m_{\rm W1}$} & \colhead{\h $\sigma_{\rm rms, W1}$} &
\colhead{\h $m_{\rm W2}$} & \colhead{\h $\sigma_{\rm rms,W2}$} &
\colhead{\h $m_{\rm W3}$} & \colhead{\h $\epsilon_{\rm W3}$} \\
\colhead{} & \colhead{(deg.)} & \colhead{(deg.)} & 
\colhead{} & \colhead{} & \colhead{} &
\colhead{(mag)} & \colhead{(mag)} & \colhead{(mag)} & \colhead{(mag)} &
\colhead{(mag)} & \colhead{(mag)} \\
\colhead{(1)} & \colhead{(2)} & \colhead{(3)} & \colhead{(4)} & \colhead{(5)} &
\colhead{(6)} & \colhead{(7)} & \colhead{(8)} & \colhead{(9)} & \colhead{(10)} &
\colhead{(11)} & \colhead{(12)}  
}
\startdata
WISE J170954.64+673101.8 & 257.4777 & 67.5172 & 1.00 & 1.00 & 0.93 & 15.49 & 0.10 & 14.24 & 0.09 & 11.21 & 0.06 \\
WISE J171011.26+672246.8 & 257.5469 & 67.3797 & 0.95 & 0.97 & 0.78 & 14.52 & 0.02 & 14.13 & 0.03 & 11.49 & 0.07 \\
WISE J171012.60+670927.1 & 257.5525 & 67.1575 & 1.00 & 1.00 & 0.92 & 15.23 & 0.07 & 14.60 & 0.11 & 11.19 & 0.06 \\
\enddata
\tablecomments{
Col. (1): Object name.
Col. (2): Right ascension (J2000).
Col. (3): Declination (J2000).
Col. (4): The probability that the sources depart from non-variability in the W1 band.
Col. (5): The probability that the sources depart from non-variability in the W2 band.
Col. (6): Correlation coefficient between the W1 and W2 bands.
Col. (7): Mean W1-band magnitude.
Col. (8): Standard deviation in the W1-band light curve.
Col. (9): Mean W2-band magnitude.
Col. (10): Standard deviation in the W2-band light curve.
Col. (11): W3 magnitude from AllWISE.
Col. (12): Error of W3 magnitude from AllWISE. \\
Only a portion of the sample is displayed here to illustrate the table structure. The comprehensive dataset is provided as supplementary electronic material.
}
\end{deluxetable*}

\begin{figure}[tp!]
\centering
\includegraphics[width=0.49\textwidth]{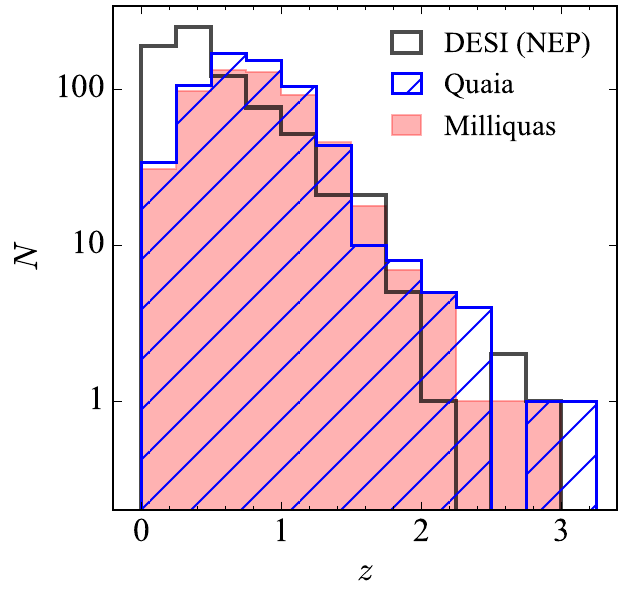}
\caption{The distributions of spectroscopic or photometric redshifts obtained from the various QSO catalogs. For objects with multiple catalog matches, we prioritize them in the following order: DESI, Quaia, and Milliquas.}
\end{figure}

Second, we cross-match our sample with the Quaia catalog (\citealp{storey-fisher_2024}), which is derived from Gaia and unWISE data (\citealp{schlafly_2019}). Quaia provides all-sky QSO candidates selected via Gaia spectrophotometric redshifts, proper motion constraints, and MIR colors. This selection method achieves high completeness, though it may result in lower purity compared to spectroscopic samples. We identify $827$ and $579$ matching sources in the NEP and SEP, respectively. We note that the Quaia catalog provides spectrophotometric redshifts based on Gaia spectra and photometric data, in combination with unWISE MIR data.  

Third, we cross-correlate our sample with the Million Quasars (Milliquas) Catalog, version 8.0 (\citealp{flesch_2023}). While this catalog is a heterogeneous compilation of sources selected across X-ray to radio wavelengths, it remains a valuable resource for identifying the origin of variability. Using a $2^{\prime\prime}$ matching radius, we find $546$ counterparts in Milliquas. We note that because Milliquas aggregates redshifts from various methods and qualities, these values should be used with caution. The distributions of redshift from the QSO catalogs are shown in Figure 5.

\subsection{Matching with Other Catalogs}
Given that the SEP region partially overlaps with the LMC, it is necessary to distinguish between Galactic or LMC stars and extragalactic sources. For this purpose, we utilize the Gaia Data Release 3 (DR3; \citealp{gaia_2023}) catalog to retrieve source proper motions. We find $2044$ and $37033$ matches from the Gaia DR3 catalog in the NEP and SEP, respectively. These proper motion measurements are included in our final catalog to facilitate further analysis.

\begin{deluxetable*}{ccccccccc}[ht!]
\tablewidth{0pt}
\scriptsize
\tablecaption{Structure Functions of Variable Sources \label{tab:tab2}}
\tablehead{
\colhead{Name} & 
\colhead{${\rm SF}^2_{0.5\ {\rm yr, W1}}$} & \colhead{${\rm SF}^2_{1\ {\rm yr, W1}}$} &
\colhead{${\rm SF}^2_{1.5\ {\rm yr, W1}}$} & \colhead{${\rm SF}^2_{2\ {\rm yr, W1}}$} & 
\colhead{${\rm SF}^2_{0.5\ {\rm yr, W2}}$} & \colhead{${\rm SF}^2_{1\ {\rm yr, W2}}$} &
\colhead{${\rm SF}^2_{1.5\ {\rm yr, W2}}$} & \colhead{${\rm SF}^2_{2\ {\rm yr, W2}}$}  \\
\colhead{} &
\colhead{(mag$^2$)} & \colhead{(mag$^2$)} & \colhead{(mag$^2$)} & \colhead{(mag$^2$)} &
\colhead{(mag$^2$)} & \colhead{(mag$^2$)} &
\colhead{(mag$^2$)} & \colhead{(mag$^2$)} \\
\colhead{(1)} & \colhead{(2)} & \colhead{(3)} & \colhead{(4)} & \colhead{(5)} &
\colhead{(6)} & \colhead{(7)} & \colhead{(8)} & \colhead{(9)}  
}
\startdata
ISE J170954.64+673101.8 & -1.78e-03 & -1.83e-04 & 1.61e-03 & 3.87e-03 & 5.60e-04 & 1.67e-03 & 2.49e-03 & 4.84e-03 \\
WISE J171011.26+672246.8 & 8.58e-07 & -2.09e-04 & -1.34e-04 & 7.60e-05 & -1.94e-04 & -6.63e-05 & -3.78e-04 & 5.77e-04 \\
WISE J171012.60+670927.1 & 1.09e-03 & 1.81e-03 & 3.02e-03 & 4.99e-03 & 1.09e-03 & 4.38e-03 & 7.39e-03 & 9.31e-03 \\
\enddata
\tablecomments{
Col. (1): Object name.
Col. (2): SF$^2$ at 0.5 yr in the W1 band. 
Col. (3): SF$^2$ at 1 yr in the W1 band. 
Col. (4): SF$^2$ at 1.5 yr in the W1 band. 
Col. (5): SF$^2$ at 2 yr in the W1 band. 
Col. (6): SF$^2$ at 0.5 yr in the W2 band. 
Col. (7): SF$^2$ at 1 yr in the W2 band. 
Col. (8): SF$^2$ at 1.5 yr in the W2 band. 
Col. (9): SF$^2$ at 2 yr in the W2 band. \\
Only a portion of the sample is displayed here to illustrate the table structure. The comprehensive dataset is provided as supplementary electronic material.
}
\end{deluxetable*}

Additionally, the relatively large point spread function (PSF; $\sim6^{\prime\prime}$) of the WISE images introduces the potential for significant source blending. To identify possible confusion, we cross-correlate our variables with the Legacy Survey Data Release 10 (LS DR10; \citealp{dey_2019}) using a $2^{\prime\prime}$ matching radius. The number of LS DR10 sources identified within this radius is recorded in our catalog as a blending flag to alert users to potentially contaminated photometry. Because LS DR10 provides only partial coverage of the SEP, the availability of supplementary information for sources within this region is limited. To address this lack of coverage, we also estimated the number of Gaia counterparts. The complete results of the cross-match with various catalogs are provided in Table 3.

Finally, to provide complementary variability information from optical time series data, we utilize the classification catalog based on ZTF light curves (\citealp{healy_2024}). \cite{healy_2024} classified variable sources within the ZTF dataset using machine learning techniques. This dataset allows us to probe the physical origins of our sources, particularly in the NEP field. Using a matching radius of 1\asec, we identify 2182 objects. Following the deep neural network (DNN) scores provided by \cite{healy_2024}, we categorize these sources into AGNs and stellar objects (including periodic and pulsating variables, binary stars, eclipsing systems, and YSOs). With a score threshold of 0.5, we find that 196 and 173 objects are classified as AGNs and stellar objects, respectively. If we instead select the classification with the maximum DNN score without imposing a threshold, 963 out of 2182 objects are identified as likely AGNs. However, classifications with relatively low scores should be interpreted with caution. These results are summarized in Table 3.    

\begin{figure*}[tp!]
\centering
\includegraphics[width=0.49\textwidth]{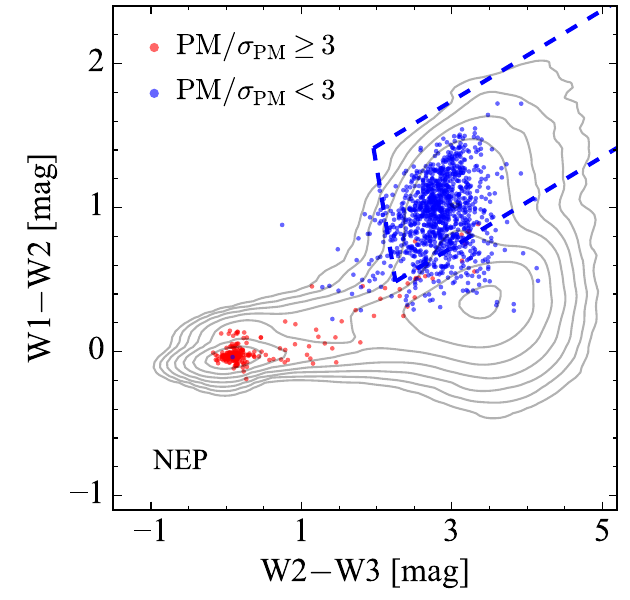}
\includegraphics[width=0.49\textwidth]{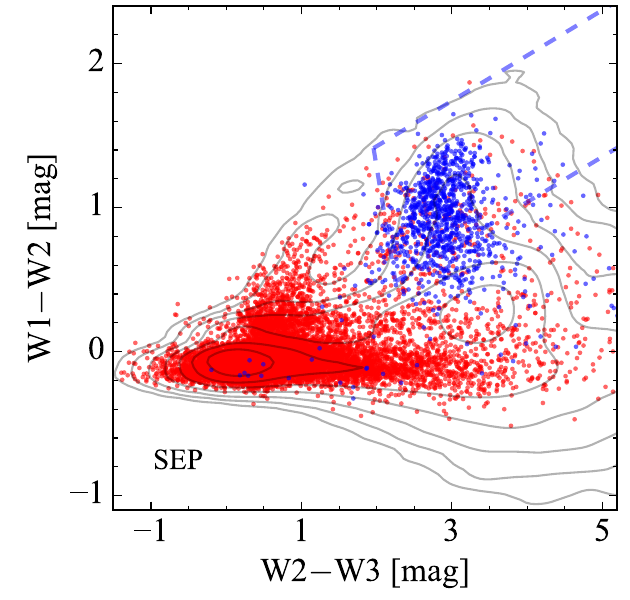}
\caption{W1$-$W2 versus W2$-$W3 MIR color diagram for the NEP (left) and SEP (right). The contours represent the distributions of the entire WISE sample. The red and blue dots indicate variable sources with and without significant proper motions, respectively. The dashed line denotes the AGN wedge adopted from \citet{mateos_2012}.}
\end{figure*}

\subsection{MIR Color-color Diagram}
The MIR color–color diagram is a powerful tool for investigating the nature of our sources, as it is sensitive to both the presence and temperature of dust \cite[e.g.,][]{lacy_2004, stern_2005, assef_2018}. Specifically, the W1$-$W2 versus W2$-$W3 color space allows for the identification of AGNs using the `AGN wedge' selection criteria \cite[e.g.,][]{mateos_2012, stern_2012, assef_2018}. We adopt W3 magnitudes from the AllWISE catalog, while W1 and W2 magnitudes are derived from the mean values of our light curves. Reliable W3 measurements (i.e., S/N $>3$) are available for $\sim80\%$ of the variables. Figure 6 illustrates the color distributions of the parent sample and variables for the NEP and SEP. In the NEP, the variables exhibit a bimodal distribution: one population resides at the locus of early-type galaxies or stars (exhibiting minimal dust contribution; \citealp{wright_2010}), while the other lies within or just below the AGN wedge, characterized by redder W2$-$W3 colors. This suggests that the majority of NEP variables are of AGN origin. More specifically, in the NEP, there are $\sim1122$ variable objects that exhibit no significant proper motion and have S/N $> 3$ in the W3 band, which can be regarded as strong candidates for AGNs. For comparison, \cite{assef_2018} identified $\sim 6414$ AGN candidates based on WISE colors in the same area of the NEP as used in this study, satisfying our selection criteria (i.e., S/N $> 10$ at W1 and W2, and S/N $> 3$ at W3). This indicates that $\sim 17\%$ of MIR-selected AGNs are variable according to our criteria. This result is broadly consistent with the findings of previous studies on MIR-variability-based AGN selection (\citealp{kims_2026}).

\begin{figure}[tp!]
\centering
\includegraphics[width=0.49\textwidth]{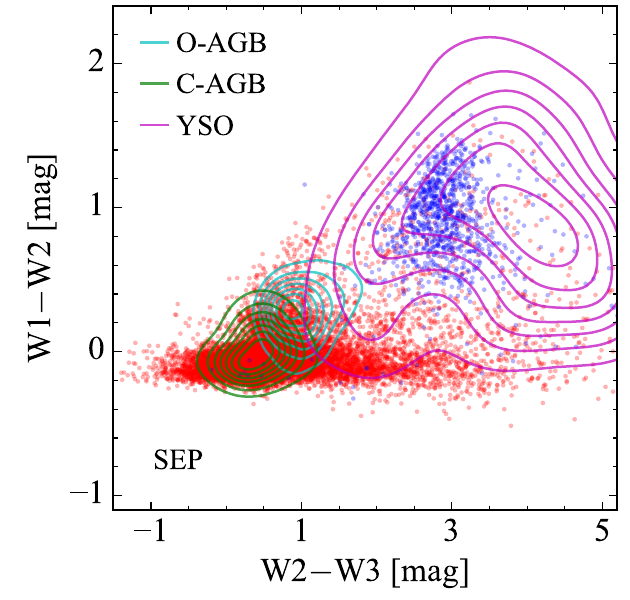}
\caption{Same as Figure 6, but with contours denoting the distributions of oxygen-rich (cyan) and carbon-rich AGB stars (green; \citealp{suh_2021}), and YSOs (magenta; \citealp{whitney_2008}).}
\end{figure}

\begin{deluxetable*}{ccccccccccc}[ht!]
\tablewidth{0pt}
\scriptsize
\tablecaption{Complementary Data\label{tab:tab3}}
\tablehead{
\colhead{Name} & 
\colhead{$z_{\rm DESI}$} &
\colhead{$z_{\rm Quaia}$} &
\colhead{$z_{\rm Mil}$} &
\colhead{Type} &
\colhead{PM} &
\colhead{$\sigma_{\rm PM}$} &
\colhead{$N_{\rm Gaia}$} &
\colhead{$N_{\rm LS}$} &
\colhead{Class$_{\rm ZTF,0.5}$} &
\colhead{Class$_{\rm ZTF}$}
\\
\colhead{} & \colhead{} & \colhead{} & \colhead{} & \colhead{} &
\colhead{(mas yr$^{-1}$)} & \colhead{(mas yr$^{-1}$)} & \colhead{} & \colhead{} &
\colhead{} & \colhead{} \\
\colhead{(1)} & \colhead{(2)} & \colhead{(3)} & \colhead{(4)} & \colhead{(5)} &
\colhead{(6)} & \colhead{(7)} & \colhead{(8)} & \colhead{(9)} & \colhead{(10)} & \colhead{(11)}  
}
\startdata
WISE J170954.64+673101.8 & \nd &1.38 & \nd & \nd & 0.621 & 0.954 & 1.0 & 1.0 & --- & E \\
WISE J171011.26+672246.8 & \nd & \nd & \nd & \nd & 0.621 & 1.523 & 1.0 & 1.0 & --- & E \\
WISE J171012.60+670927.1 & \nd & \nd & \nd & \nd & 0.344 & 2.932 & 1.0 & 1.0 & --- & Q \\
\enddata
\tablecomments{
Col. (1): Object name.
Col. (2): Redshift from DESI spectra. 
Col. (3): Redshift from the Quaia catalog. 
Col. (4): Redshift from the Milliquas catalog. 
Col. (5): DESI spectral classification: G (galaxies), Q (QSOs), or S (stars). 
Col. (6): Proper motion from Gaia DR3. 
Col. (7): Measurement errors for proper motions. 
Col. (8): Number of Gaia counterparts within a matching radius of 2 \asec. 
Col. (9): Number of Legacy Survey counterparts within a matching radius of 2 \asec. \\
Col. (10): Classification of variable sources based on ZTF light curves (\citealp{healy_2024}), using a deep neural network (DNN) score threshold of $>0.5$: Q (QSOs), E (eclipses), S (variable stars), P (pulsating sources), B (binary stars), Y (YSOs).
Coll. (11): Classification of variable sources. Categories are assigned based on the maximum DNN score provided by \cite{healy_2024} without a minimum selection criterion. \\
Only a portion of the sample is displayed here to illustrate the table structure. The comprehensive dataset is provided as supplementary electronic material.
}
\end{deluxetable*}

Conversely, SEP variables are more broadly distributed across the color–color space, indicating that stellar sources are the dominant contributors to variability in this field. To refine our source classifications, we divide the sample based on the Gaia proper motion (${\rm PM}$) significance. We define sources with significant proper motion as those with ${\rm PM}/\sigma_{\rm PM} > 3$. In the NEP, these significant-${\rm PM}$ sources primarily occupy the stellar locus. In the SEP, they are located outside the AGN wedge. These results confirm that the combination of proper motion and MIR colors effectively distinguishes the physical origins of the observed variability. 

\begin{figure*}[tp!]
\centering
\includegraphics[width=0.95\textwidth]{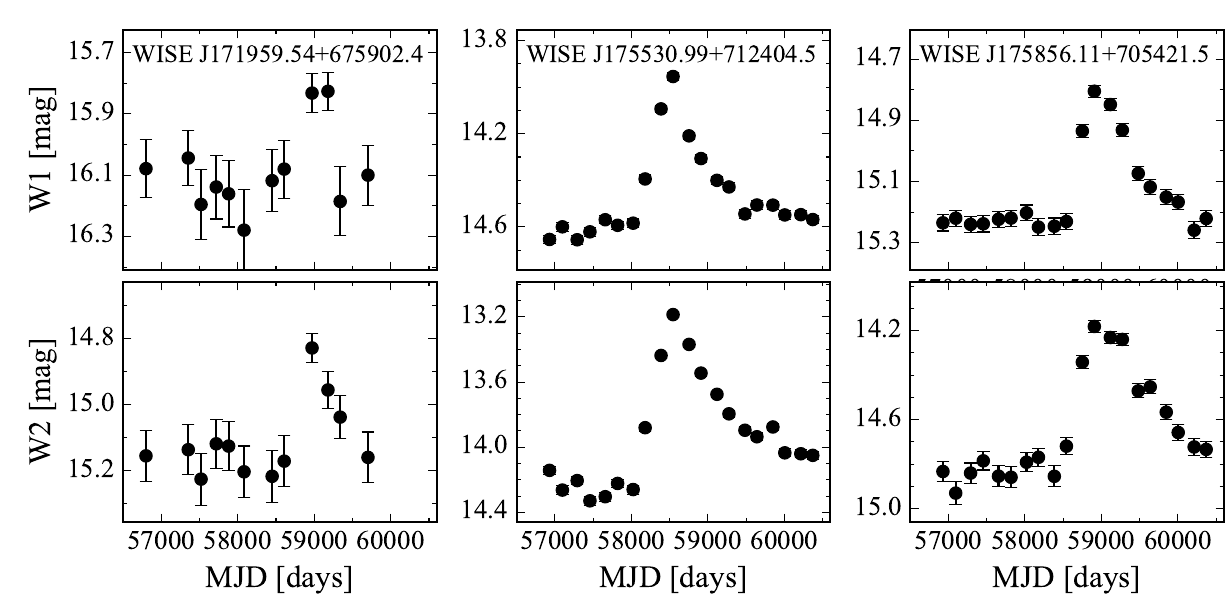}
\caption{MIR light curves of the three identified variable sources exhibiting transient behavior.}
\end{figure*}

To further assess the origin of the variable sources in the SEP, we compare their colors with those of evolved stars, such as asymptotic giant branch (AGB) stars from \cite{suh_2021}, and YSOs in the LMC from \cite{whitney_2008}, both of which are known to be luminous and variable in the MIR. As shown in Figure 7, the locations of these evolved stars in the MIR color-color diagram coincide with those of MIR variable sources with significant PMs. This finding further confirms that the majority of variable sources in the SEP are attributable to stellar activity.

\section{MIR Transients}
As an immediate application of our dataset, we attempt to identify transient objects, which are essential for studying supernovae or TDEs in dust-enshrouded regions. We conduct this experiment solely on the NEP, where complementary optical data are available to further investigate the physical origins of these phenomena. Through visual inspection, we identify only three objects exhibiting a sudden increase in brightness followed by a gradual decrease (light curves shown in Figure 8). To examine their physical origin, we modeled the spectral energy distribution (SED), covering optical ($grizy$ from Pan-STARRS; \citealp{chambers_2016}) and MIR photometric data, using the LePHARE code \citep{arnouts_1999, ilbert_2006}. For the SED fitting, we adopt templates for inactive galaxies from the COSMOS survey \citep{ilbert_2009}, obscured and unobscured AGNs \citep{lyu_2017, byun_2023, 2024ApJS..275...46K}, and stars \citep{bohlin_1995, pickles_1998}.

We find that all objects are well-fit by obscured AGN templates with moderate photometric redshifts ranging from $z=0.07$ to $0.60$ (Fig. 9). This suggests that the MIR transients are likely caused by nuclear activity, such as a TDE or a nuclear flare, although a supernova origin remains possible considering the enhanced SF within the AGN host galaxies \citep{zhuang_2021}. Although the sample size is small, it is interesting to note that these transients are detected only in obscured AGNs, suggesting a physical association between nuclear transients and circumnuclear obscuration. \cite{son_2022b} found an MIR-only transient in NGC 3786 (an intermediate-type AGN associated with a merging feature) and suggested it originated from a TDE or AGN flare occurring in a dust-rich nucleus. Notably, those results are consistent with our findings, supporting the idea that such MIR transients may occur preferentially in obscured AGNs.        

\section{Applications}
The SPHEREx mission conducts a spectrophotometric all-sky survey using linear variable filters (LVFs), providing low-resolution optical-to-MIR spectral coverage over $0.75\text{--}5.0\mu\rm{m}$. The survey depth and observing strategy of SPHEREx are comparable to those of the WISE mission. Due to its wide field of view, SPHEREx continuously observes the ecliptic poles, allowing for the acquisition of multi-epoch spectrophotometric data similar to those obtained by WISE \cite[e.g.,][]{kim_2021jkas}. However, the incoherent cadence of spectral coverage inherent to the LVF-based observing strategy complicates the identification of variable sources using SPHEREx data alone \cite[e.g.,][]{bryan_2025}. As a result, the variables identified in this study provide valuable reference samples for SPHEREx-based investigations. Moreover, the spectrophotometric capabilities of SPHEREx will enable detailed studies of variable sources; for example, MIR spectral variability in AGNs can place meaningful constraints on the dust covering factor \cite[e.g.,][]{son_2022,son_2023}. In addition, SPHEREx will significantly extend the temporal baselines of MIR light curves, which is essential for characterizing sources with long variability timescales, such as luminous AGNs \cite[e.g.,][]{kim_2024}.

\begin{figure}[tp!]
\centering
\includegraphics[width=0.49\textwidth]{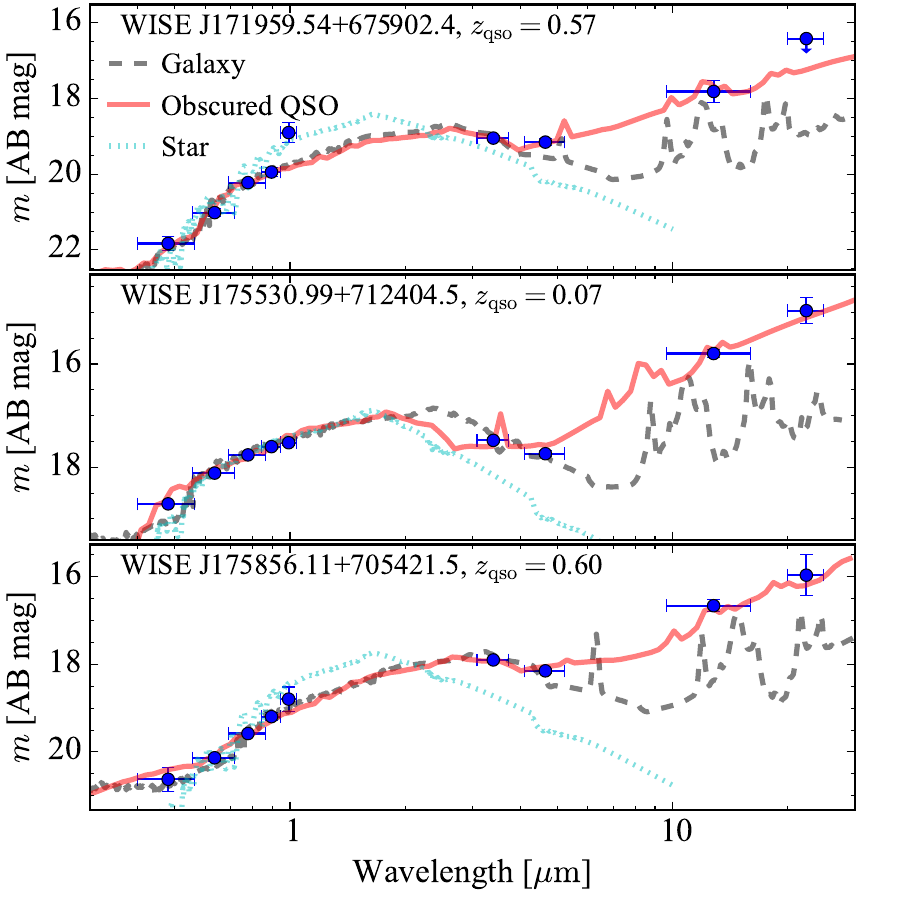}
\caption{SED fitting results for the transient targets. Blue circles represent the observed photometry, while black dashed, red solid, and cyan dotted lines denote the best-fit templates for inactive galaxies, QSOs, and stars, respectively. All targets are well-fit by the obscured QSO templates. }
\end{figure}

Several optical time-domain surveys, including Zwicky Transient Facility (ZTF; \citealp{bellm_2019}), LSST, and 7DS, provide continuous coverage of the ecliptic poles. Comparisons between MIR and optical light curves offer valuable insights into the physical properties of a wide range of astrophysical sources. For example, reverberation mapping studies of AGNs use the time lag between optical and MIR variability as a tracer of the dusty torus size, requiring quasi-simultaneous observations in combination with the WISE dataset. Such analyses have been conducted by pairing existing optical light curves with all-sky WISE observations, for which the typical cadence is $\sim6$ months \cite[e.g.,][]{lyu_2019, yang_2020}.  The enhanced sampling of the ecliptic poles relative to other regions of the sky is particularly advantageous for robust lag measurements. However, reverberation mapping is only feasible when the optical and MIR observations are temporally well aligned, limiting the applicability of future optical surveys for this purpose. Instead, comparative analyses of variability characteristics in optical and MIR light curves, without explicitly measuring time lags, can still provide meaningful constraints on the physical properties of AGN dusty tori \cite[e.g.,][]{kim_2024, son_2026}. Consequently, the combination of ongoing and forthcoming optical surveys, such as 7DS and LSST, with MIR datasets will remain highly valuable.

\section{Summary}
Taking advantage of the frequent visits of the WISE mission to the ecliptic poles and its complementarity to other surveys, we construct a catalog of MIR variables within a circular area with a 5-degree radius at the north and south ecliptic poles. To ensure reliable photometric measurements and minimize outliers, we carefully processed the data, including a correction for focal-plane temperature variations that cause marginal shifts in the zero-point. For a robust estimation of photometric uncertainties, we recalculated the errors using the entire WISE sample at the ecliptic poles. To facilitate the precise identification of variables, we employ two parameters: $P_{\rm var}$, which denotes the probability that the observed light curves are intrinsically variable, and $r$, the correlation coefficient between W1 and W2 magnitudes. We additionally discarded objects exhibiting variability caused by the influence of neighboring bright sources. This selection process results in 2702 variables in the NEP and 27514 variables in the SEP; the proximity of the Large Magellanic Cloud (LMC) significantly increases the number of variables in the SEP.       

By performing a systematic cross-match with existing QSO catalogs, we determine that at least $\sim6.5\%$ of our detected variables are likely to be AGNs. This high-confidence subset offers a valuable foundation for subsequent statistical analyses of AGN variability. Furthermore, by integrating high-precision proper motion measurements from the Gaia DR3 catalog with MIR color-color diagnostics, we distinguish a distinct population of variables of stellar origin. The contribution of these stellar sources is significantly more pronounced in the SEP region, a result directly attributable to the proximity of the LMC and its high density of evolved stars and young stellar objects. 

As an immediate application of the dataset, we identify three MIR transient objects in the NEP region. These transients, characterized by a sudden increase in brightness followed by a gradual decline, are all well-fit by obscured AGN templates at moderate redshifts between $z = 0.07$ and $0.60$. The findings suggest that these events likely originate from nuclear activity, such as TDEs or nuclear flares, while a supernova origin cannot be completely ruled out.  Notably, the detection of these transients exclusively in obscured AGNs aligns with previous research, supporting the hypothesis that nuclear transients are physically associated with dust-rich circumnuclear environments.

Ultimately, the catalog presented here is designed to serve as a baseline for the next generation of time-domain astronomy. When utilized in conjunction with upcoming multi-epoch datasets from facilities such as the 7DS, the LSST, SPHEREx, and the ZTF, our sample will provide a unique opportunity to disentangle various physical mechanisms. This synergy will enable us to more clearly elucidate the physical origins of MIR variability, whether driven by stochastic accretion disk fluctuations in AGNs, the pulsating atmospheres of evolved stars, or young stellar objects.

\begin{acknowledgments}
We are grateful to the anonymous referee for the very constructive review that has improved our manuscript greatly.
L.C.H. was supported by the National Science Foundation of China (12233001) and the China Manned Space Program (CMS-CSST-2025-A09). S.S. was supported by the KIAA and Boya Fellow Grant of Peking University. This work was supported by the National Research Foundation of Korea (NRF) grant funded by the Korean government (MSIT) (Nos. RS-2024-00347548 and RS-2025-16066624) and the Yonsei University Research Fund of 2025 (2025-22-0402). B.L. is supported by the NRF grant funded by the Korean government (MSIT) (No. 2022R1C1C1008695).

\end{acknowledgments}

\facilities{WISE, NEOWISE}

\software{astropy \citep{2013A&A...558A..33A,2018AJ....156..123A,2022ApJ...935..167A}, }

\bibliography{ms}{}
\bibliographystyle{aasjournalv7}

\appendix
\section{Example Light Curves of High Cadence}

Figure~A1 presents example high-cadence light curves.
\restartappendixnumbering

\begin{figure*}[htp]
\centering
\includegraphics[width=0.97\textwidth]{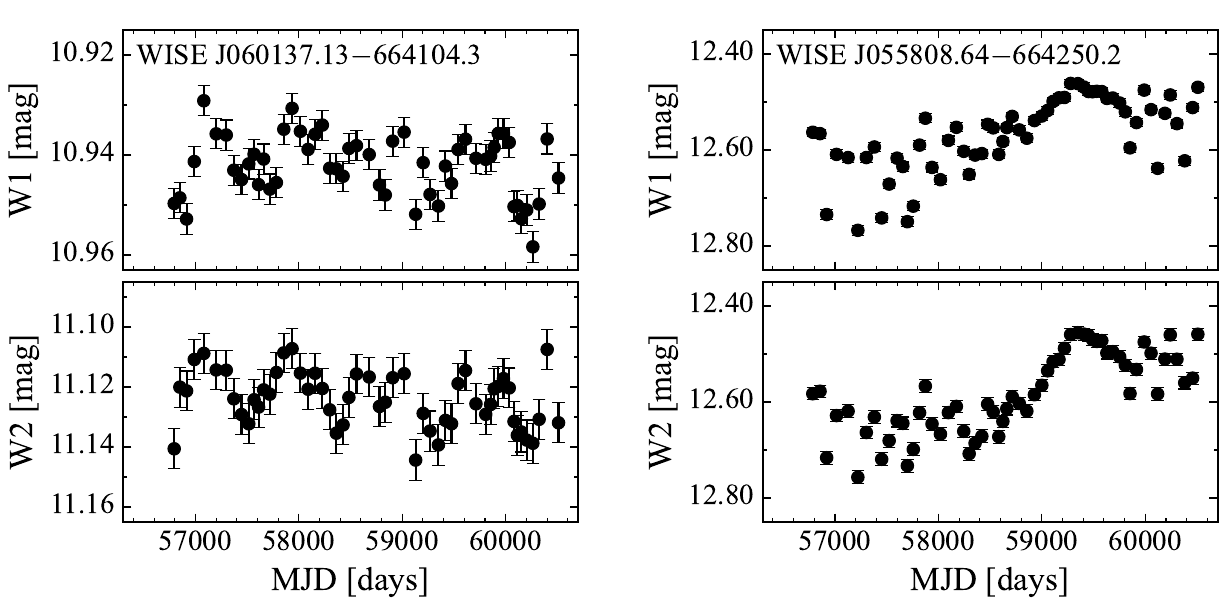}
\caption{Examples of W1 (top) and W2 (bottom) light curves observed with high cadences.}
\end{figure*}

\end{document}